\RequirePackage{etex} 
\documentclass[aps,prb,english, longbibliography, onecollum, 11pt]{revtex4-2}
\usepackage[active]{srcltx}
\usepackage[utf8]{inputenc}
\usepackage{latexsym}
\usepackage[hidelinks]{hyperref}
\usepackage{amsmath}
\usepackage{xcolor}
\usepackage{amssymb}
\usepackage{amsfonts}
\usepackage{graphicx}
\usepackage{slashed}
\usepackage{footmisc}
\DeclareSymbolFont{usualmathcal}{OMS}{cmsy}{m}{n}
\DeclareSymbolFontAlphabet{\mathcal}{usualmathcal}
\usepackage{braket}

\newcommand{\be}{\begin{equation}}
	\newcommand{\ee}{\end{equation}}
\newcommand{\bea}{\begin{eqnarray}}
	\newcommand{\eea}{\end{eqnarray}}

\usepackage{microtype}
\usepackage{physics}
\usepackage{indentfirst}
\usepackage{dcolumn}
\usepackage{bm}

 \makeatletter 
 \patchcmd\Gread@eps{\@inputcheck#1 }{\@inputcheck"#1"\relax}{}{}
 \makeatother

\begin{document}
\title{Boundary and Symmetry Breaking in a Deformed Toric Code}
\author{Rodrigo Corso}
\affiliation{Departamento de Física, Universidade Estadual de Londrina\\Caixa Postal 10011, 86057-970, Londrina, PR, Brasil.}
\email{rodrigocorso@uel.br}
\begin{abstract}
	This work explores a deformation of Kitaev’s toric code that induces a phase transition out of the topologically ordered phase. By placing the model on a cylinder, the bulk global 1-form symmetries separate into distinct boundary operators, allowing us to show that the transition is accompanied by the condensation of all anyons in a given boundary. Using a lower dimensional $(1+1)$-dimensional boundary Hamiltonian, we extract an effective central charge and find a pronounced suppression near $\beta_c$, followed by its restoration at strong coupling.
\end{abstract}

\maketitle

\section{Introduction}
Many strongly correlated quantum systems exhibit topological order, a form of organization not captured by any local order parameter. The prime example is the fractional quantum Hall, whose states have long-range entanglement and a ground-state degeneracy that cannot be lifted by any local perturbation \cite{wen1990ground}. Such topologically ordered phases may host exotic fractionalized excitations and are intrinsically robust to local noise, a property that has motivated proposals for decoherence-resistant quantum memories and fault-tolerant quantum computation \cite{wen2013topological,kitaev2003fault,moore1991nonabelions}. By its very nature, these phases lie beyond the Landau paradigm: they are characterized by nonlocal entanglement and global properties of the ground state wave function rather than local symmetry-breaking \cite{nussinov2009symmetry}. 

Unlike 0-form symmetry-broken phases described by the Landau-Ginzburg-Wilson framework, topologically ordered phases are best understood in terms of their entanglement structure \cite{wen2013topological,xu2023entanglement}. A hallmark of such order is a finite, universal correction to the  entanglement entropy, the topological entanglement entropy, which captures the presence of long-range quantum correlations in the ground state \cite{PhysRevLett.96.110405, kitaev2006topological}. In gauge-theoretic realizations of topological order, different phases can be distinguished by the expectation values of non-contractible Wilson and 't Hooft loops: these follow a perimeter law in the deconfined (topological) phase and an area law in the confined (trivial) phase. These loop operators are the natural generators of higher-form symmetries \cite{Gaiotto2015}, generalized symmetries defined by extended objects rather than point-like fields. In two spatial dimensions, the mutual 't Hooft anomaly between a pair of 1-form symmetries underlies the topoligical order, directly leading to the ground-state degeneracy on nontrivial manifolds. This perspective, developed in recent years, unifies gauge-theoretic, categorical, and condensed-matter descriptions of topological phases \cite{Moradi2023, McGreevy_2023, bhardwaj2023categorical, thorngren2024fusion, Lichtman2021}. 

An archetypal example is Kitaev’s toric code model, an exactly solvable spin Hamiltonian on a two-dimensional lattice \cite{kitaev2003fault}. The toric code realizes a $\mathbb{Z}_{2}$ topologically ordered phase. Its ground state degeneracy depends on the genus of its manifold, as it supports two species of anyonic quasiparticles (often called ``electric''  and ``magnetic" charges) with mutual anticommutation. Crucially, these anyons and the ground-state degeneracy are protected by the model's topological symmetries structure and are robust against local perturbations. Such property forms the basis for topological quantum error correction and fault-tolerant quantum computation. As such, the toric code serves as both a pedagogical prototype and a building block for more intricate topological systems, from string-net models to stabilizer codes and Majorana-based qubits.

Physical realizations of the toric code have been actively pursued across several experimental platforms. Trapped-ion and superconducting qubit architectures were among the first to implement its stabilizer structure at small scales, demonstrating the creation and manipulation of anyonic excitations \cite{barreiro2011open,corcoles2015demonstration}. One direct realization was achieved by Google Quantum AI, who implemented the surface code, an open-boundary variant of the toric code with the same $\mathbb{Z}_{2}$ topological order, on a superconducting processor and demonstrated error correction below the fault-tolerance threshold \cite{satzinger2021realizing,google2023suppressing}. In parallel, Verresen et al. predicted that a topological quantum liquid with toric code topological order could be realized in a two-dimensional Rydberg atom array on the ruby lattice \cite{verresen2021prediction}, a prediction subsequently confirmed experimentally \cite{semeghini2021probing}. These developments raise the natural question of how robust the topological order is under perturbations, in particular, whether the deformation studied here leaves detectable signatures in boundary observables and entanglement scaling accessible to these platforms.

A natural question is how this long-range entanglement can be dissolved as one drives the system into a trivial phase.  A well-studied example is the deformation introduced in \cite{castelnovo2008quantum}, which is controlled by the parameter $\beta$. For $\beta=0$, one recovers the Kitaev toric code, while as $\beta$ increases the state continuously approaches a magnetized product state. Remarkably, this model remains exactly solvable for all $\beta$. Its analysis shows that the topological entanglement entropy stays at the toric-code value until a critical $\beta_c$, where it abruptly drops to zero \cite{castelnovo2008quantum}, a signal that the system undergoes a phase transition from a topologically ordered phase to a trivial phase. This model has been used to study anyon condensation and confinement, as $\beta$ increases certain anyons condense and the phase loses its topological order \cite{Huxford:2023bhb}.

The deformed toric code model presents a particularly interesting example of a topological-trivial phase transition. The transition at $\beta_c$ is sharply signaled by the abrupt vanishing of the topological entanglement entropy \cite{castelnovo2008quantum}, which directly measures the long-range entanglement structure underpinning topological order. Remarkably, however, not all topological properties are simultaneously destroyed. The ground state degeneracy, symmetries of the Hamiltonian and anyon content remain invariant across the transition, persisting into the trivial phase. This decoupling between topological entanglement entropy and ground state degeneracy reveals a more intricate structure than a naive picture of topological order destruction, and raises the question of which observables faithfully diagnose the transition.

Topological phases of matter are not only distinguished by their bulk properties, but also exhibit remarkable physics at their boundaries. The prototypical example is the fractional quantum Hall effect, whose chiral edge modes are described by chiral CFTs fixed by the bulk Chern-Simons theory \cite{wen1992chiral,Moore1989,Ardonne2004}. More broadly, string-net models provide a lattice realization of Turaev-Viro TQFTs, and their gapped boundary theories have been systematically classified in terms of the input fusion category \cite{kitaev2012models,levin2005string}.

The toric code is a lattice realization of a Turaev-Viro TQFT built from the fusion category $\mathrm{Rep}(\mathbb{Z}_2)$ \cite{levin2005string}. In this construction, gapped boundaries arise via the condensation of bulk anyons and are classified by Lagrangian algebras of the anyon model \cite{kitaev2012models}. By fine tuning a boundary parameter, the boundary undergoes a phase transition between such gapped phases at a gapless critical point, described by a CFT whose data reflects the bulk topological order. In the toric code, this fine-tuned critical boundary realizes the Ising CFT with $c=1/2$, directly reflecting the $\mathbb{Z}_2$ anyon structure of the bulk.

In this light, we study the $\beta$-deformed toric code on an open manifold. By placing the model on a cylinder, the two bulk non-contractible loop operators split into a closed symmetry and an open string operator connecting the two boundaries. We show that the topological-to-trivial phase transition at $\beta_c$ is directly tied to a degradation of the symmetry structure of the model. This degradation is cemented through the non-trivial behavior of an order parameter built from correlators of open electric lines, computed via a population Monte Carlo method exploiting the mapping to the classical 2D Ising model.

Motivated by the connection between the toric code and the Ising CFT, we derive a $(1+1)$-dimensional boundary Hamiltonian from the bulk and show that it is precisely identified with the axial next-nearest-neighbor Ising (ANNNI) model. This identification yields a critical line for the boundary, departing from the naive transverse-field Ising result due to the non-perturbative role a $ZZ$ interaction introduced by the $\beta$ deformation.

Finally, we compute the effective central charge of the boundary theory along this critical line using DMRG simulations for chains up to $L=500$ sites, by fitting the Calabrese-Cardy entanglement entropy scaling formula. We find that the central charge exhibits a striking non-monotonic dependence on $\beta$: it is strongly suppressed in the vicinity of the bulk critical point $\beta_c$ and subsequently restored to the undeformed Ising value $c=1/2$ at strong coupling. We interpret this suppression as a boundary signature of bulk gaplessness near $\beta_c$, consistent with the recently proposed intermediate gapless regime \cite{Verressen2025}, rather than as a direct indicator of bulk topological order.

This work is organized as follows. Section \ref{model} presents the model of study and discusses its main features and points of interest. Section \ref{topological phases} considers the model in a cylinder and examines the boundary condensing phase structure that arises when we place the model in an open manifold. Section \ref{symmetry breaking} approaches the phase transition from the spontaneous symmetry breaking of a bulk, global 1-form symmetry. At last, in Section \ref{holography} we discuss the holographic correspondence between the bulk and the boundary.

\section{The Model}\label{model}
\begin{figure}[h]
	\centering
	\includegraphics[width=0.35\linewidth]{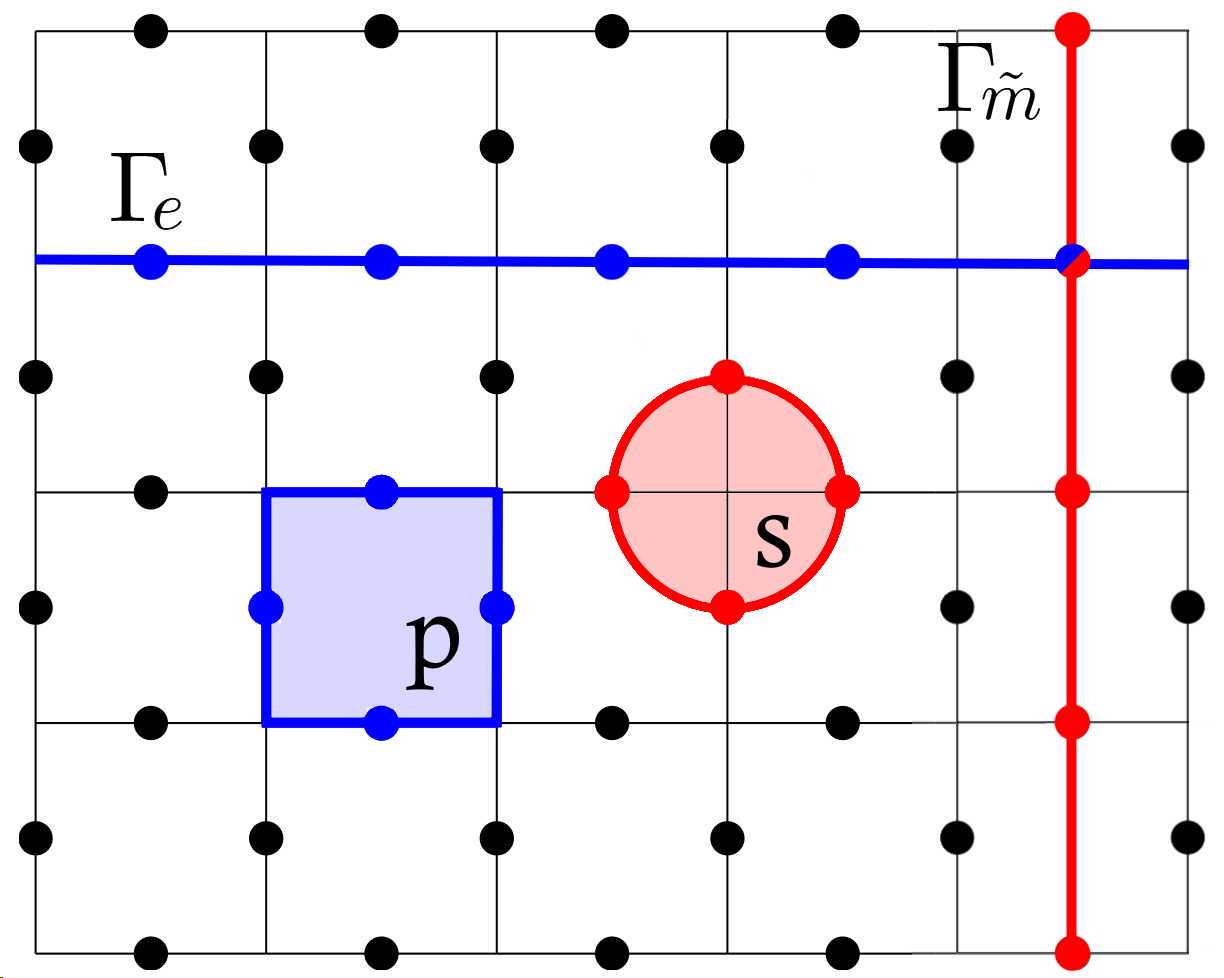}
	\caption{Electric and magnetic degrees of freedom on the toric code, color online.}
	\label{fig:lattice2}
\end{figure}

In this section we review some of the important aspects of our particular deformation of interest \cite{castelnovo2008quantum}. Like the standard toric code \cite{kitaev2003fault}, the deformation can be defined in any graph $G$. For simplicity we restrict ourselves to the simplest case of a square lattice, where on each link we place a spin-1/2 variable, see Fig.\ref{fig:lattice2}. For this review section, we will initially assume periodic boundary conditions (PBC), but later on this will be relaxed to study the effect of the deformation on open boundaries. The Hamiltonian for the deformed toric code (DTC) is given by
\begin{align}
	H_{blk}=\mathcal J_{p}\sum_{p}(1-B_{p})+\mathcal J_{s}\sum_{s}e^{-\frac{\beta}{2}\sum\limits_{\vec r\in s} Z_{\vec r}}(1-A_{s})e^{-\frac{\beta}{2}\sum\limits_{i\in s}Z_{\vec r}}, \label{Hamiltonian}
\end{align}
where $\vec{r}=(i,j)$ localizes a spin variable in the two dimensional lattice. Furthermore, the operators $B_{p}=\prod_{\vec r\in p} Z_{\vec r} $ and $A_{s}=\prod_{\vec r\in s}X_{\vec r}$ are defined as the product of Pauli matrices $Z$ and $X$ over edges of plaquettes (p) and stars (s), as in Fig. \ref{fig:lattice2}. One may notice that the deformation only affects the star operators, since it trivially commutes with the plaquette operators.

The Hamiltonian \eqref{Hamiltonian} is gapped for generic $\beta$. For $\beta < \beta_c$, the gap is protected by topological order and is robust against local perturbations \cite{Hastings2004,Ardonne2004,Hastings2010}. At the critical coupling $\beta_c$, the gap closes and the system undergoes a topological to trivial phase transition \cite{castelnovo2008quantum}. For large $\beta$, the model enters a trivially gapped confined phase, although it has recently been argued that an intermediate gapless regime may persist over a finite window above $\beta_c$ \cite{Verressen2025}, a point we return to in Section \ref{holography}.

In fact, the addition of the deformation spoils the commuting projector structure of the original model, since $A_s$ does not commute with the deformation, and the deformed star operator
\begin{align}
	H_s(\beta)&=e^{-\frac{\beta}{2}
		\sum\limits_{\vec r\in s} Z_{\vec r}}(1-A_{s})e^{-\frac{\beta}{2}\sum\limits_{\vec r\in s}Z_{\vec r}}.
\end{align}
is not a projector. Nonetheless, we are still able to find the ground states in a systematic way. The deformed star operators are positive semi-definite as a result of 
\begin{align}
	H_s^2(\beta) &= \cosh\left(\beta\sum_{\vec r\in s}Z_{\vec r}\right)H_s(\beta).
\end{align}
From which we immediately see that the eigenvalues of $H_{s}$ must be positive, and therefore, it suffices to look for a state that is annihilated by all the $H_s$ operators. This is accomplished by the following set of states
\begin{equation}
	\ket{GS(\beta)}\propto S(\beta) \prod_{s} (1+A_{s}) \ket{\Uparrow}
	\label{eq:gs_state}
\end{equation}
where $\ket{\Uparrow}$ is the fully magnetized state and the deformation $S(\beta)=\exp\left((\beta/2)\sum_{\vec r}Z_{\vec r}\right)$. In fact, the ground states in \eqref{eq:gs_state} belong to a more general class of wavefunctions known as \textit{square root states} \cite{McGreevy2016}, which were first observed in the context of a generalization of Rokhsar-Kivelson Hamiltonians and connection to classical Markovian dynamics \cite{Castelnovo2005}. 

Nonetheless, the ground state \eqref{eq:gs_state} is not the only ground state of the model. In fact, for periodic boundary conditions the DTC Hamiltonian hosts two higher-form, non-contractible loop symmetries. The electric symmetry is defined by the operator
\begin{align}
	 \Gamma_{e,\gamma}=\prod_{\vec r\in \gamma} Z_{\vec r}
\end{align}
where $\gamma$ is  closed path winding around one of the handles of the torus, running along the links of the lattice,  see Fig. \ref{fig:lattice2}. This is an ordinary symmetry of the model, an unitary operator that commutes with the Hamiltonian \eqref{Hamiltonian}, such that it is a symmetry of the complete Hilbert space. Additionally, the model also hosts a magnetic symmetry
\begin{align}
	\tilde \Gamma_{m,\gamma^{\star}}=S(\beta)\left[\prod_{\vec r\in \gamma^{\star}} X_{\vec r}\right] S(\beta)^{-1},
\end{align}
where $\gamma^{\star}$ is a closed path along one the handles of the torus running across the links, or equivalently on the dual lattice. This symmetry only commutes with the Hamiltonian when acting on the ground states, such that it is only a symmetry of the ground states subspace. Furthermore, in general it is not a unitary operator, a subject we discuss shortly.

As the symmetry operators along different directions on the torus do not commute,
\begin{align}
	\Gamma_{e,\gamma}\tilde\Gamma_{m,\gamma^{\star}}=(-1)^{\text{link}(\gamma,\gamma^{\star})}\tilde \Gamma_{m, \gamma^\star} \Gamma_{e,\gamma}, \label{algebra}
\end{align}
the ground states cannot be simultaneously diagonalized by both symmetry operators. This scheme is often referred to as a 't Hooft anomaly and leads to the topological degeneracy of the model \cite{gomes2023introduction}. On the torus, there are four topologically distinct ground states
\begin{align}
	\ket{GS;\omega_{x},\omega_y}=\frac{1}{\sqrt{\mathcal Z_{\omega_x,\omega_y}}}S(\beta)(\Gamma_{m,x})^{\omega_x} (\Gamma_{m,y})^{\omega_y}\sum_s (1+A_s)\ket{\Uparrow}, \label{gs1}
\end{align}
where $\omega_{x,y}=0,1$ indicates the absence or presence of a non-contractible loop in the $x$ or $y$ direction along the torus. The normalization constant $\mathcal{Z}_{\omega_x,\omega_{y}}$ is equivalent to the two dimensional classical Ising partition function, where the parameters $\omega_{x}$ and $\omega_y$ represent a $\mathbb{Z}_{2}$ twist along the respective directions on the torus. At last, the operator
\begin{align}
	\Gamma_{m,x}=\prod_{\vec r \in \gamma^{\star}_{x}}X_{\vec{r}} \label{ge}
\end{align}
is the non-deformed (undressed) magnetic non-contractible loop in the $ x $ direction of the torus. 

These ground states are graded by the action of the electric symmetries
\begin{align}
	\Gamma_{e,x}\ket{GS;\omega_x,\omega_y}=(-1)^{\omega_y}\ket{GS;\omega_x,\omega_y}\;\;\;\; \text{and}\;\;\;\;	\Gamma_{e,y}\ket{GS;\omega_x,\omega_y}=(-1)^{\omega_x}\ket{GS;\omega_x,\omega_y},
\end{align}
and the magnetic symmetry is spontaneously broken, such that it maps between the topologically distinct ground states
\begin{align}
	\ket{GS;\omega_x,\omega_y}= \left(\frac{\mathcal{Z}_{0,0}}{\mathcal{Z}_{\omega_{x},0}}\right)^{\omega_{x}/2} \left(\frac{\mathcal{Z}_{0,0}}{\mathcal{Z}_{0,\omega_{y}}}\right)^{\omega_{y}/2}(\tilde \Gamma_{ m, x})^{\omega_x} (\tilde \Gamma_{m,y})^{\omega_y}\ket{GS; 0,0}. \label{norm_factors}
\end{align}

We can get some intuition from these ground states by recasting them in the $\mathbb{Z}_2$ gauge theory language \cite{Wegner1971, Kogut1975, Kogut1979,Wilson1974}. The ground state of $H_{blk}$ for $\beta=0$ can be simply interpreted as the equal weight superposition of all contractible loops generated by the action of the product of star operators $A_s$. Then as $S(\beta)$ acts on $\ket{GS}$, it produces a different weight to each state in the superposition. Since the presence of a loop makes the eigenvalues of some $Z$-operators flip to $-1$, the weight of the exponential factor is decreased for large loop configurations.

In this way, the $\beta$ deformation introduces tension for the loop strings, which has the effect of suppressing large loop configurations, leading to a confined phase. From this analysis, we can already preclude that the model goes through a phase transition at some critical value $\beta_c$. For $\beta<\beta_c$, the string tension of the loops is not large, and we sit at a deconfined phase (topological order), while for $\beta>\beta_c$ large loops start to be suppressed, leading to a confined phase, culminating in a trivial product state in the $\beta\rightarrow \infty$ limit. Recently, the authors in \cite{Verressen2025} defied this picture, arguing that there may be a gapless regime in between the confined/deconfined phases. We will soon comment more on this subtle point. This effect is analogous to the loop gas picture studied in \cite{PhysRevLett.98.070602}, where a loop tension drives a continuous phase transition to a magnetic phase. 

Indeed, the DTC is known to undergo a continuous phase transition at $\beta=\beta_{c}\equiv \frac{1}{2}\ln \left(1+\sqrt{2}\right)$ \cite{castelnovo2008quantum}. This was first unveiled as the topological entanglement entropy suddenly jumps from the toric code topological order value, to zero. This presents a curious kind of phase transition, from a topologically ordered phase to a topologically trivial phase, whithout explicitly breaking any symmetry of the torus Hamiltonian.

In this representation, the normalization factors $ \mathcal Z_{\omega_{x},\omega_y} $ can be interpreted in a simple fashion. They are the partition functions of the classical Ising model in two spatial dimensions at inverse temperature $\beta$. This is made more explicitly once we make the identification $Z_{\vec r} = \theta_{\vec r-1/2}\theta_{\vec r+1/2}$, where $\theta_{\vec r\pm 1/2}$  are classical Ising variables located at the vertices bounding the spin $Z_{\vec r}$. 

This mapping leads to an important relation between the quantum DTC and the classical Ising model. The action of a single star operator $A_s$ on the sum \eqref{gs1} can be encoded on the Ising model by flipping the Ising variable on the same vertex. Then, the sum over all loop configurations on the toric code can be encoded by a sum over Ising configurations. A detailed discussion on this connection can be found in \cite{castelnovo2008quantum, Huxford:2023bhb}. This equivalence enables the calculation of the expected values of two types of operators. First, for general functions of $Z$
\begin{align}
	\expval{f(Z_{\vec r})}_{DTC}=\expval{f(\theta_{\vec r-1/2}\theta_{\vec r+1/2})}_{Ising}. \label{ising mapping 1}
\end{align}
Additionally, a similar relation holds for the product of $ X_{\vec r} $ along a closed path
\begin{align}
	\langle\prod_{\vec r\in \gamma^{*}}X_{\vec r}\rangle_{DTC}
	&=\frac{1}{2^{N/2}\sqrt{\mathcal Z_{0,0}}}\bra{0,0}\left(\prod_{\vec r\in \gamma^*}e^{-\beta Z_{\vec r}}\right)S(\beta)\left(\prod_{\vec r\in \gamma^*}X_{\vec r}\right)\prod_s (1+A_s)\ket{\Uparrow}_{DTC}\nonumber\\
	&=\langle \prod_{\vec r\in \gamma^*} e^{-\beta Z_{\vec{r}}}\rangle_{DTC}
	\Rightarrow \langle e^{-\beta\sum\limits_{\vec r\in \gamma^{*}} \theta_{\vec r-1/2}\theta_{\vec r+1/2}} \rangle_{Ising}.  \label{ising mapping 2}
\end{align}

\begin{figure}[t]
	\centering
	\includegraphics[width=0.45\linewidth]{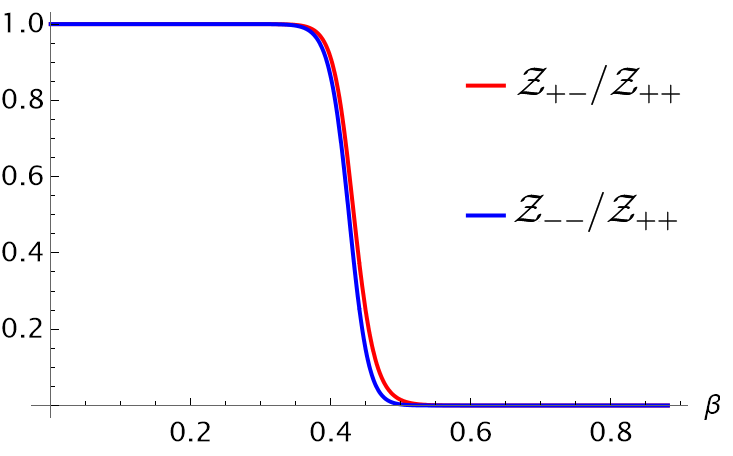}
	\caption{Ratio of the normalizations against the coupling $ \beta $.}
	\label{fig:partition-function-ratios}
\end{figure}

This connection between the these two models was instrumental in the detection of the critical point in \cite{castelnovo2008quantum}. Furthermore, it also allows us to effectively compute expected values in the DTC ground states by mean values in the classical Ising model at inverse temperature $\beta$. While obtaining exact analytical results in the Ising model may be challenging, these quantities may be efficiently estimated using well-established Monte Carlo methods, an approach we adopt in subsequent discussions.

We close this section with a remark on the 't Hooft anomaly and the phase structure of the model. The operator algebra \eqref{algebra} is an exact operator identity that holds for all $\beta$. The anomaly it encodes is a UV constraint: it forbids a trivially gapped symmetric phase in which both $\Gamma_e$ and $\tilde\Gamma_m$ are simultaneously unbroken and the spectrum is non-degenerate. 

The DTC gives different solutions to this issue depending on the value of $\beta$. For $\beta<\beta_c$, the anomaly leads to the spontaneous breaking of $\tilde{\Gamma}_{m}$, which maps between the four degenerate ground states on the torus, this is the very origin of the topological degeneracy. At $\beta=\beta_c$, the gap closes and the anomaly is matched by the gapless critical point \cite{castelnovo2008quantum,Hastings2004,Ardonne2004}. 

For $\beta>\beta_c$, the reconciliation is more subtle. The mapping between the ground states \eqref{norm_factors} indicates that the ratio between the Ising partition functions serves as a measure of the unitarity of $\tilde \Gamma_m$ \cite{Huxford:2023bhb}. As shown in Fig. \ref{fig:partition-function-ratios}, such ratio rapidly vanishes above $\beta_c$. Even though \eqref{algebra} continues to hold as an operator identity, $\tilde\Gamma_m$ becomes non-unitary and it progressively loses its status as a legitimate symmetry. The anomaly structure is not violated but degraded: as $\tilde\Gamma_m$ ceases to be a proper symmetry, the constraint no longer prevents a trivially gapped confined phase. We develop this picture further in Section \ref{symmetry breaking}.


\section{Open Boundary}\label{topological phases}

\begin{figure}[h]
	\centering
	\includegraphics[width=0.6\linewidth]{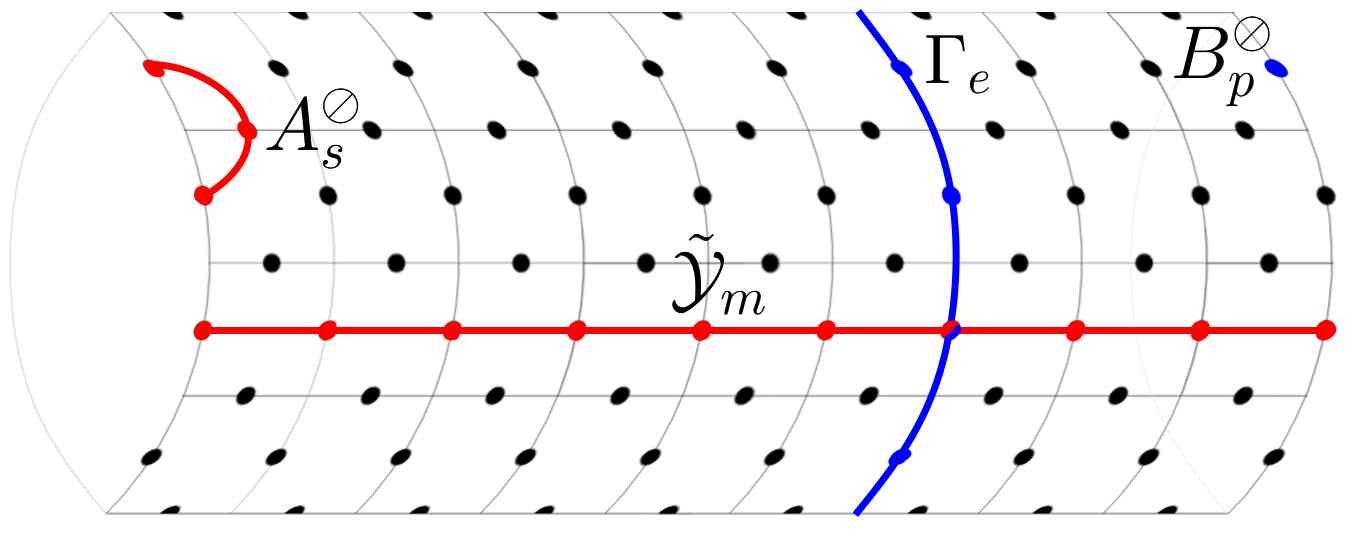}
	\caption{Smooth boundary on the cylinder. $ B_{p}^{\oslash} $ should be thought at just one site.}
	\label{fig:cilinder-lattice-smooth-lattice}
\end{figure}

Topological order in two spatial dimensions manifests not only in the bulk but also through the way the system is terminated. Cutting open one of the non-contractible cycles of the torus introduces decoupled degrees of freedom which, if left ungapped, generate a ground-state degeneracy that scales extensively with system size. To preserve the toric code phase, it is necessary to include in the Hamiltonian boundary operators that gap out these edge degrees of freedom while commuting with the bulk Hamiltonian \eqref{Hamiltonian} on the ground-state manifold.

A natural choice for such terms are the incomplete versions of the star and plaquette operators. For the standard toric code ($\beta=0$), these are the $ A_{s}^{\oslash}=\prod_{\vec r \in s^{\oslash}} X_{\vec r} $ and $ B_{p}^{\oslash}=Z $ operators illustrated in Fig. \ref{fig:cilinder-lattice-smooth-lattice}. As these boundary operators do not commute between each other, they generate two competing boundary phases in the toric code topological ordering. These are characterized by the condensation to the ground state of open strings terminating on the boundary, or equivalently, by the spontaneous breaking of the associated global 1-form symmetries.

A natural question arises in how this condensation and symmetry breaking  scheme is affected by the introduction of the deformation $\beta $. To this end, we consider our model on a cylinder, obtained by cutting open one of the directions of the torus. We perform a cut along the $y$ direction of the torus, such that this coordinate maintains its periodicity, as in Fig. \ref{fig:cilinder-lattice-smooth-lattice}.

In doing so, the $\Gamma$ operators running along the different directions in the torus split. As before, we reserve the label $\Gamma$ for the operators running along the periodic direction of the torus
\begin{align}
	\Gamma_{e}=\prod_{\vec r \in \gamma_{y}} Z_{\vec{r}}\qquad\qquad\text{and} \qquad \qquad \tilde \Gamma_{m}=S(\beta)\left[\prod_{\vec r\in \gamma^{\star}_{y}}X_{\vec{r}}\right]S(\beta)^{-1},
\end{align}
where $\gamma_{y}$ and $\gamma^{\star}_{y}$ are non-contractible closed paths winding along the periodic direction of the cylinder on the direct and dual lattices, respectively. Upon cutting open the torus, the operators running along the $x$ direction acquire endpoints on the two boundaries. We distinguish these open lines from their closed counterparts by labeling them
\begin{align}
	\mathcal{Y}_{e}=\prod_{\vec{r} \in \gamma_{x}}Z_{\vec{r}} \qquad\qquad\text{and}\qquad\qquad \mathcal{\tilde Y}_{m}=S(\beta)\left[\prod_{\vec{r}\in \gamma^{\star}_{x}} X_{\vec{r}}\right]S(\beta)^{-1},
\end{align}
where $\gamma_{x}$ and $\gamma_{x}^{\star}$ are open paths along the $x$ direction of the cylinder on the direct and dual lattices, respectively. In therms of the new operators, the operator algebra \eqref{algebra} reads
\begin{align}
	\mathcal{Y}_{e}\tilde\Gamma_{m}=-\tilde\Gamma_{m}\mathcal{Y}_{e} \qquad\qquad\text{and} \qquad \qquad \mathcal{\tilde Y}_m\Gamma_{e}=-\Gamma_{e}\mathcal{\tilde Y}_{m}. \label{open_algebra}
\end{align}

As the algebra leads to two different anyon condensing phases on the boundary of the toric code, the anticommutation relations \eqref{open_algebra} imply that no boundary Hamiltonian can simultaneously gap the edge degrees of freedom while preserving both pairs of operators as symmetries. Instead, one must choose: either magnetic open lines terminate trivially at the boundary, condensing magnetic anyons, leaving $\Gamma_e$ spontaneously broken, or electric open lines $\mathcal{Y}_e$ condense and $\tilde{\Gamma}_m$ is broken. These two choices define two competing gapped boundary phases in the toric code topological order.

The specific choice of which excitations condense is fixed by adding to the bulk Hamiltonian one of two boundary terms, each gapping the edge while commuting with the bulk on the ground-state manifold. The deformation modifies this picture since both the condensate and the symmetry structure acquire a coupling dependence. To make this explicit, we now examine each fixed-point boundary Hamiltonian in turn.

The first Hamiltonian we discuss is produced by the inclusion of the incomplete star operators, $ A_{s}^{\oslash} $, accompanied by the deformation
\begin{align}
	H_{\partial \mathcal{M}}^{M}=\sum_{s\in \partial \mathcal{M}}e^{-\frac{\beta}{2}\sum\limits_{\vec r\in s} Z_{\vec r}}\left(1-A_{s}^{\oslash}\right)e^{-\frac{\beta}{2}\sum\limits_{\vec r\in s} Z_{\vec r}}. \label{boundary 1}
\end{align}
Such inclusion alters the symmetry content of the model. The incomplete star operator does not commute with $\mathcal{Y}_{e}$, such that it is not a symmetry of $H_{blk}+H^{M}_{\partial \mathcal{M}}$. Conversely, $\Gamma_{e}$ does commute with the Hamiltonian, and $\mathcal{\tilde Y}_{m}$ and $\tilde \Gamma_{m}$ commute when applied to the ground state. The 't Hooft anomaly between $\mathcal{\tilde Y}_{m}$ and $\Gamma_{e}$ still holds, such that the ground state cannot simultaneously diagonalize both symmetries. We set a basis that diagonalizes $\mathcal{\tilde Y}_{m}$
\begin{align}
	\mathcal{\tilde Y}_{m}\ket{\pm}_M=\pm\ket{\pm}_{M},
\end{align} 
and spontaneously breaks $\Gamma_{e}$
\begin{align}
	\Gamma_{e}\ket{\pm}_{M}=\sqrt{\frac{Z_{\pm}}{Z_{\mp}}} \ket{\mp}_M, \label{mag mapping}
\end{align}
where $Z_{\pm}$ are normalization constants. This is achieved by the set of states
\begin{align}
	\ket{\pm}_{M}=\frac{1}{\sqrt{Z_{\pm}}}S(\beta)(1\pm\mathcal{Y}_{m})\prod_{s\in\partial\mathcal{M}}(1+A_{s}^{\oslash})\prod_{s\in\mathcal{M}}(1+A_{s})\ket{\Uparrow}\label{gs magnetic},
\end{align}
where $\mathcal{Y}_{m}$ is equal to $\mathcal{\tilde Y}_{m}$ for $\beta=0$.

The $\beta$ deformation affects the condensation scheme for the magnetic boundary. The dressed magnetic line on the boundary $ \tilde A_{s}^{\oslash}=S(\beta) A_s^{\oslash} S^{-1}(\beta) $ remains constant for all $\beta$
\begin{align}
	\expval{\tilde A_s^{\oslash}}_{M}=1. \label{expval1}
\end{align}
Conversely, the electric line 
\begin{align}
    \expval{B_{p}^{\oslash}}_{M}=\begin{cases}
	0\text{ for }\beta=0\\
	f_{e}(\beta)\text{ for }\beta\neq0
\end{cases} \label{19},
\end{align}
acquires some non-trivial dependence for $\beta>0$.

In general, states obtained by the action of open lines are excited states. For the boundary fixed point $ H_{\partial \mathcal M}^{M} $, when the endpoints of a deformed magnetic line are pushed to the boundary, the line operator may be rewritten as a product of complete $\tilde A_s=S(\beta) A_s S^{-1}(\beta) $ and incomplete $\tilde A_s^{\oslash}$ deformed star operators. In a gauge theoretic language, this boundary also includes lines ending on the boundary on its ground state superposition of loops. 

As the ground state \eqref{gs magnetic} is an eigenstate of deformed star operators, the magnetic excitations condense to the ground state, if their endpoints are pushed to the boundary. This is readily seen from \label{expaval1} by depicting \eqref{expval1} as the superposition between the open line state $ \tilde A_s^{\oslash}\ket{\pm}_M $ and the ground state $ \ket{\pm}_M $. 

Furthermore, the $M$ condensation trivializes the action of $ \tilde \Gamma_m $, as it can be written as the product over all $ \tilde A_s^{\oslash} $. Conversely, $ \Gamma_{e} $ cannot be absorbed by the $ M $ condensate, such that it remains a non-trivial symmetry in the low energy limit. 

We may also produce an electric fixed point Hamiltonian by including the incomplete plaquette operators $ B_p^{\oslash}=Z_i $, as in Fig. \ref{fig:cilinder-lattice-smooth-lattice}
\begin{align}
	H_{\partial \mathcal{M}}^{E}=\sum_{p\in \partial \mathcal{M}}\left(1-B_p^{\oslash}\right) \label{boundary 2}.
\end{align}
For such case, the degenerate ground states include only the bulk star operators
\begin{align}
	\ket{\omega}_{E}=\frac{1}{\sqrt{\mathcal Z_{\omega}}}e^{\frac{\beta}{2}\sum\limits_{\vec r\in \mathcal{M}_{blk}}Z_{\vec r}}\left(\Gamma_{m}\right)^{\omega}\prod_{s\in \mathcal{M}_{blk}}(1+A_s)\ket{\Uparrow},
\end{align}
where $\omega=0,1$ is a $\text{mod }2$ variable, $Z_{\omega}$ is a normalization constant, which we still identify with the partition function for the Ising model, and $\Gamma_{m}$ is $\tilde{\Gamma}_{m}$ for $\beta=0$.

 These ground states are graded by the electric symmetry 
\begin{align}
	 \mathcal{Y}_{e}\ket{\omega}_{E}=(-1)^{\omega}\ket{\omega}_{E},
\end{align}
and spontaneously break the magnetic symmetry
\begin{align}
	\tilde \Gamma_{m}\ket{\omega}_{E}=\sqrt{\frac{Z_{\omega+1}}{Z_{\omega}}}\ket{\omega}_{E}. \label{elec mapping}
\end{align}
Furthermore, the action of $\Gamma_{e}$ is trivial: $ \Gamma_e\ket{\omega}_{E}=\ket{\omega}_E $.

For this boundary Hamiltonian, the expected value of the boundary operators remain constant for all $\beta$
\begin{align}
	\expval{\tilde A_{s}^{\oslash}}_{E}=0,\;\;\;\;\text{and}\;\;\;\; \expval{B_p^{\oslash}}_{E}=1. \label{electric condensation}
\end{align}
The electric boundary Hamiltonian condenses open electric lines with endpoints in a boundary. This can easily seen by observing that these lines can be written as a product complete and incomplete plaquette operators. In contrast, open magnetic lines lead to excited states. 

The two boundary Hamiltonians do not commute, but they do commute with the bulk Hamiltonian. In this way, they compete to form two different boundary phases in the toric code topological order, such that we study phase transitions via a general Hamiltonian
\begin{align}
	H=H_{blk}+J_{p}H_{\partial\mathcal{M}}^{E}+J_{s}H_{\partial\mathcal{M}}^{M}. \label{eqq2}
\end{align}

\section{Anyon Condensation} \label{symmetry breaking}

\begin{figure}[t]
	\centering
	\includegraphics[width=0.6\linewidth]{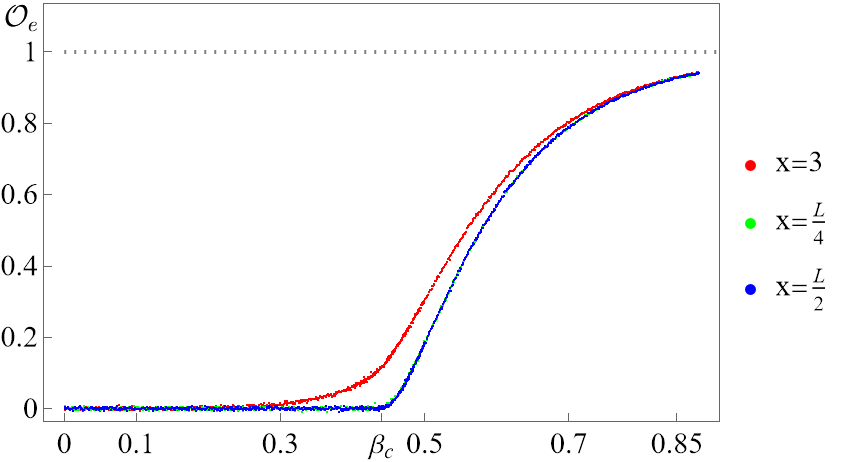}
	\caption{Order parameter $ \mathcal{O}_{e}(\beta) $ for $ L=64 $ on a square lattice for the $ \ket{0} $ ground state with varying line distance. Notice that the green and blue points are almost coincident.}
	\label{fig:orderpar}
\end{figure}

The toric code $(\beta=0)$ phase transition is associated with the spontaneous breaking of its \textit{global} 1-form symmetries. At the $J_p=0$ fixed point, $\tilde \Gamma_{m}$ is a trivial symmetry, the pair $\Gamma_{e}$, $\mathcal{\tilde{Y}}_{m}$ is in a symmetry breaking scheme, and $\mathcal{Y}_{e}$ leads to excitations. In the previous section, we presented a basis where $\Gamma_{e}$ is broken and $\mathcal{\tilde Y}_{m}$ is intact, acting as an order parameter for $\Gamma_{e}$. The symmetry roles are reversed as we cross the $J_s=J_p$ phase transition. At the $J_s=0$ fixed point, $\Gamma_{e}$ is trivial and the pair $\tilde \Gamma_{m}$, $\mathcal{Y}_{e}$ is in a 't Hooft anomaly. This scheme guarantees that in either phase, there is always a pair $\Gamma$, $\mathcal{ Y}$ that is in an spontaneous breaking scheme, one $\Gamma$ is trivial and one $\mathcal{Y}$ leads to excitations. Combined with the electromagnetic duality, this symmetry structure is fundamental in the realization of topological phases and their long range entanglement \cite{xu2023entanglement,zhao2021higher,hirono2024symmetry,barkeshli2024higher,Yoshitome:2025klz,Moradi2023}.

In general, this 1-form symmetry breaking picture is accompanied by the condensation of boundary anyons. Let us take the magnetic boundary as an example. This fixed point is achieved by the inclusion of magnetic boundary operators to the Hamiltonian, thus, magnetic open lines ending on the boundary commute with the Hamiltonian and condense to the ground state. This includes $\mathcal{\tilde{Y}}_{m}$, an open line the normally would lead to an excitation, that is present in the ground state manifold.

Turning on the $\beta$ deformation greatly affects the anyon condensation content of the model. Firstly, the $\beta$ deformation mostly affects the condensation scheme of the electric line on the magnetic boundary, see equations \eqref{19} and \eqref{electric condensation}, so we focus on this boundary. Furthermore, the magnetic symmetry on the torus becomes non-unitary as the ratio between the normalizations vanishes for $\beta>\beta_c$, see Fig. \ref{fig:partition-function-ratios}.

In this light, we study the condensation of electric anyons on the magnetic boundary through the study of the order parameter
\begin{align}
	\mathcal{O}_{e}=\expval{ \mathcal{ Y}_{e}(x) \mathcal{ Y}_{e}(0)}_{M}, \label{order_par}
\end{align} 
which signals the condensation of electric anyons whenever $\mathcal O _{e}\neq 0$ for large separation $x$ \cite{PhysRevX.3.021009,Lichtman2021}. Note $\mathcal{Y}_{e}$ is not a symmetry of the magnetic boundary Hamiltonian.

\begin{figure}[t]
	\centering
		\includegraphics[width=0.6\linewidth]{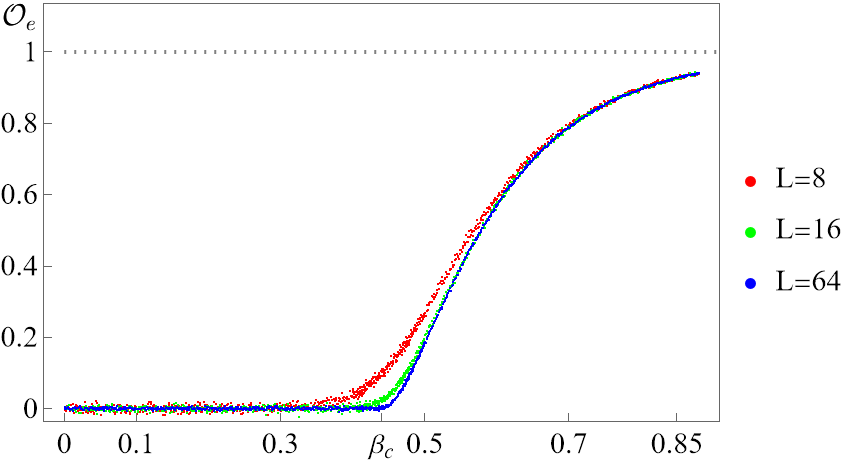}
	\caption{Order parameter at fixed separation $x=L/2$ and varying system size $L=8,16,64$ for a square lattice.}
	\label{fig:orderparvarl}
\end{figure}

Directly computing this order parameter from the deformed toric code ground state is a challenging task. For this reason, we are urged to make use of the mapping between expected values in the deformed toric code and thermal mean values in the Ising model described in equations \eqref{ising mapping 1} and \eqref{ising mapping 2}. With this connection at hand, we are able to effectively simulate the order parameter using a populational Monte Carlo method \cite{weigel2021understanding,machta2010population}. This method is specially well suited to our applications as it considers the normalized contributions from the most important loops in the ground state sum to the order parameter.

The resulting estimator for the order parameter can be found in Fig. \ref{fig:orderpar} for a fixed system sized, with varying line separation $\mathbf{x}$. Notice that the latter two lines, for $\mathbf{x}=L/4$ and $\mathbf{x}=L/2$, are almost coincident, confirming that the order parameter is independent of line separation for large $\mathbf{x}$. We also present a plot for $\mathcal{O}_{e}$ with maximum line separation and varying system size $L$ in Fig. \ref{fig:orderparvarl}, confirming that the simulations achieve a satisfactory thermodynamic limit approximation.

 Above the critical coupling, $\beta_c\approx 0.44$, $\mathcal{O}_e\neq 0$ indicating that, for the magnetic boundary, the phase transition is second order and accompanied by the condensation of electric open lines terminating on the boundary. 
 
 Note that this anyon condensation picture is not backed by the SSB of $\tilde{\Gamma}_{m}$, as it remains trivial for all $\beta$. Even though the algebra
 \begin{align}
 	\tilde{\Gamma}_{m}\mathcal{Y}_{e}=-\mathcal{Y}_{e}\tilde{\Gamma}_{m}
 \end{align}
remains a true statement between operators, $\mathcal{Y}_{e}$ is not a symmetry and does not produce a 't Hooft anomaly with $\tilde \Gamma_{m}$, which remains trivial. Curiously, this does produce non-trivial effects on $\mathcal{\tilde Y}_{m}$ when we choose a basis where it is spontaneously broken, like in the electric boundary. As we discussed for the periodic boundary, the ration between the normalization constants in the mapping between the ground states \eqref{elec mapping} acquires some non-trivial dependence above $\beta_{c}$, indicating that $\mathcal{\tilde Y}_{m}$ acts non-unitarily on the ground state.

This is fundamentally different from the usual toric code electric-magnetic phase transition. In the non-deformed model, the two phases are connected by a 0-form symmetry. Such that, as we cross the self-dual $J_s=J_p$ critical point the roles played by the electric and magnetic symmetry are simply reversed. This process does not change the global symmetry structure of the toric code, such that we may simply exchange the labels $e \Leftrightarrow m$.

The symmetry structure consisting of two mutually ’t Hooft anomalous 1-form symmetries, tied together by a global electromagnetic 0-form symmetry, stabilizes the toric code topological order and underlies many of its defining properties \cite{Moradi2023, McGreevy_2023}. In particular, the ground state degeneracy on nontrivial manifolds is a direct manifestation of the ’t Hooft anomaly between electric and magnetic line operators \cite{Gaiotto2015, gomes2023introduction}, while the nontrivial quantum dimensions of the anyons account for the finite topological entanglement entropy \cite{PhysRevLett.96.110405, kitaev2006topological}.

Upon crossing the critical point $\beta_c$, two parallel changes occur: on the magnetic boundary, electric anyons condense in addition to the already-present magnetic condensate, so that both species of anyons are condensed on the same boundary. Simultaneously on the electric boundary, the magnetic symmetry $\mathcal{\tilde Y}_{m}$ becomes non-unitary, as signaled by the ratio in \eqref{elec mapping}. Since the electromagnetic duality exchanges $e \leftrightarrow m$, it would normally map one boundary phase into the other. However, above $\beta_c$ the two boundaries are no longer symmetric: the anyon content of the electric boundary remains unchanged while the magnetic boundary hosts a full condensate of both anyon species. The two boundary phases are therefore no longer related by electromagnetic duality above $\beta_c$.

We argue that the condensation of both $e$ and $m$ anyons on the magnetic boundary above $\beta_{c}$ (while the electric remains unchanged), the non-unitarity of the 1-form magnetic symmetry $\mathcal{\tilde Y}_{m}$ and the breakdown of the 0-form electromagnetic duality, reflect a significant degradation of the symmetry structure of the model and are the primary mechanism driving the destruction of toric code topological order, as reflected in the vanishing of the topological entanglement entropy for $\beta > \beta_c$.

\section{Boundary Theory}\label{holography}

In this section, we investigate the bulk topological phase transition from the perspective of a holographic boundary theory. For both boundaries, $\Gamma_{d}$ are topological symmetry operators of the Hamiltonian. As such, we may transport this line to act only on the boundary degrees of freedom by the action of bulk plaquette and star operators. We may construct a holographic (lower dimensional) Hamiltonian by considering only the spins of the complete (2+1) dimensional Hamiltonian \eqref{eqq2} located at the boundary. As $ \Gamma_{d} $ is topological and the product of local operators, it is also a global symmetry of the holographic Hamiltonian. In this way, we investigate the bulk higher-form symmetry breaking phase transition using a lower dimensional edge theory as a proxy \cite{Lichtman2021, thorngren2024fusion, PhysRevB.108.075105, Moradi2023}.

The edge Hamiltonian is obtained by taking only the spins positioned along one of boundaries of the Hamiltonian \eqref{eqq2}, this yields the holographic, (1+1) dimensional Hamiltonian
\begin{align}
	H_{edg}&=J_{p}\sum_{i}(1-Z_{i})+J_{s}\sum_{i}\left(e^{-\beta(Z_{i}+Z_{i+1})}-X_{i}X_{i+1}\right),\nonumber\\
	&\propto -(y+\sinh 2\beta)\sum_{i} Z_{i}-\sum_{i} X_{i}X_{i+1}\nonumber\\
    &+\sinh^{2}(\beta)\sum_{i}Z_{i}Z_{i+1}. \label{boundary hamiltonian},
\end{align}
where $ y\equiv J_p/J_s $.

In general, a bulk (2+1)-dimensional topological order admits gapless degrees of freedom on its boundary, often being described by a CFT \cite{Moore1989, Moradi2023,moore1991nonabelions, many_body}. In the usual toric code, $\beta=0$, the holographic Hamiltonian is the transverse field Ising model, which realizes the  $c=1/2$ Ising CFT by tuning $y=1$ \footnote{Remember that $y$ is a boundary parameter  and the bulk topological ordering should not depend on a specific choice of $y$}. In such case, as we cross the $y=1$ critical point the model undergoes a topological phase transition from the toric code's electric to the magnetic condensing phases. Likewise, its topological boundary transitions from a ferromagnetic to a paramagnetic phase. 

\begin{figure}[t]
	\centering
	\includegraphics[width=0.6\linewidth]{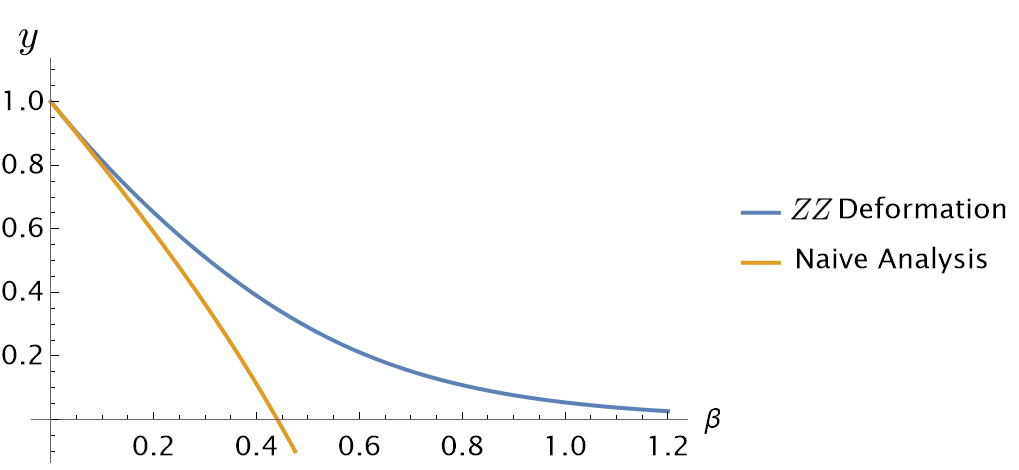}
	\caption{Critical lines with and without the $ Z_{i}Z_{i+1} $ term.}
	\label{fig:phasediag}
\end{figure}

A first, naive, analysis might assume the $ Z_{i}Z_{i+1} $ interaction to be irrelevant in the thermodynamic limit. Under a Jordan-Wigner transformation, such term maps to $ -\gamma_{i}\gamma_{i}^{\prime}\gamma_{i+1}\gamma_{i+1}^{\prime} $, which is irrelevant in the low-energy limit by a simple power counting. These considerations would reduce our analysis to the transverse field Ising model with a different parametrization. The boundary would undergo a phase transition whenever the coefficients of the $ Z_{i} $ and $ X_i X_{i+1} $ terms are equal:
\begin{align}
	y=1-\sinh 2\beta, \label{naive crit}
\end{align}
which is the yellow plot in Fig. \ref{fig:phasediag}. At this critical line the model would host a Kramers-Wannier symmetry, which is inherited from the bulk electromagnetic duality.

The bulk electromagnetic duality constrains the theory even when such  marginal operators are neglected, such that this naive analysis reveals the bulk critical point. The boundary parameter $ y $ is restricted to be positive as the bulk requires $ J_{p} $ and $ J_s $ to be positive, to produce the correct boundary loop condensations. For $ \beta>\beta_c\approx 0.441 $ there is no setting of the boundary parameter $ y $ that fully hosts the bulk electromagnetic symmetry, since the bulk is in an electric condensing phase. 

In spite of the naive analysis, the $ Z_iZ_{i+1} $ deformation is known to induce non-trivial effects \cite{slagle2021microscopic,sela2011orbital,katsura2015exact, hassler2012strongly}. Firstly, such term may be seen as a dynamical mechanism for mass generation by giving an expected value for $ Z_{i} $. This displaces the critical line away from the naive analysis \eqref{naive crit}. Beyond this perturbative argument, the full holographic Hamiltonian admits a non-perturbative identification with the next-nearest-neighbor Ising (ANNNI) model, for which the ferromagnetic-paramagnetic phase transition is known to be at 
\begin{align}
	1-\frac{2U}{\Delta}=\frac{1}{\Delta}-\frac{U}{2\Delta^{2}\left(\Delta-U\right)}, \label{crit_line}
\end{align} 
for $\Delta\equiv y+\sinh 2\beta$ and $U\equiv \sinh^{2}\beta$. The critical line in terms of the variables most natural to our problem $y(\beta)$ is depicted in blue in Fig. \ref{fig:phasediag}.

For small $\beta$ the $ZZ$ deformation is still small and we may tune $y$ such that the $ c=1/2 $ Ising criticality survives, indicating that the bulk toric code topological order is also present for small $\beta$. Conversely, for greater values of $\beta$, the $ZZ$ deformation produces more subtle effects. At the critical line, we may recast our model as a $ T \bar T $ like deformation of the transverse field Ising model at criticality, which are known to realize higher central charge CFTs \cite{rahmani2015phase, datta2018t, griguolo2022phase, leclair2021t, slagle2021microscopic}. 

Let us consider a Hamiltonian of the general form
\begin{align}
	H=H_{Ising}+\lambda H_{def}.
\end{align} 
The $\lambda$ deformation must be Kramers-Wannier invariant in order to not induce a mass gap in the model. For large enough $\lambda$ this deformation leads to a CFT that contains the Ising symmetries, leading to a higher central charge. There are two deformations that are known to be UV complete \cite{leclair2021t}, the first is achieved by setting
\begin{align}
	H_{def}=\sum_{i}X_{i-1}X_{i}Z_{i+1}+Z_{i-1}X_{i}X_{i+1},\label{sym_def_2}
\end{align}
which leads to the tri-critical Ising CFT with $c=7/10$, for strong $\lambda$. Furthermore, we may also set
\begin{align}
	H_{def}=\sum_{i}Z_{i}Z_{i+1}+X_{i-1}X_{i+1}, \label{sym_def_1}
\end{align}
which leads to a $c=3/2$ CFT for sufficiently large $\lambda$.

In contrast to the Hamiltonians \eqref{sym_def_2} and \eqref{sym_def_1}, the $ZZ$ term alone found in the holographic Hamiltonian explicitly breaks Kramers-Wannier self duality, a hallmark of the Ising criticality that must be included in a UV CFT. Hence, for $\beta $ sufficiently high at the red critical line in Fig. \ref{fig:phasediag}, the holographic Hamiltonian may not be conformally invariant.

To investigate this possibility, we compute the holographic model's central charge as a function of beta along the critical line \eqref{crit_line}. To do so, we obtain the holographic Hamiltonian ground state by performing a DMRG simulation using the standard ITensor library \cite{fishman2022itensor} for varying values of $\beta$ and system size. For the largest, $L=500$, the simulation's truncation error was kept $\leq 10^{-10}$ and after 30 sweeps the ground state energy converged to at least 11 significant digits. 

\begin{figure}[t]
	\centering
	\includegraphics[width=1.\linewidth]{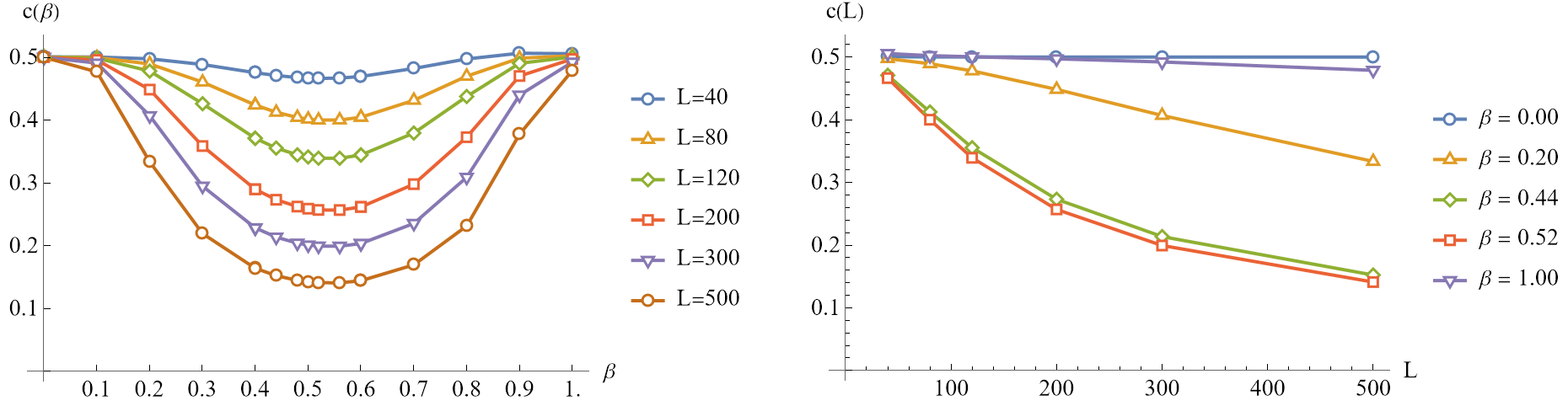}
	\caption{Plots for $c(\beta)$ for fixed $L$ (left), and $c(L)$ for fixed $\beta$ (right) obtained by fitting the Calabrese-Cardy formula.}
	\label{fig:centralcharge}
\end{figure}

From the DMRG ground states, we compute the von Neumann entanglement entropy $S(\ell) = -\mathrm{Tr}\,\rho_\ell \log \rho_\ell$ for all subsystem sizes $\ell = 2, \cdots, L/2$, where the upper limit follows from the $S(\ell)=S(L-\ell)$ symmetry of the periodic chain. We then fit this set of values to the Calabrese--Cardy scaling formula for systems with periodic 
boundary conditions \cite{calabrese2004entanglement},
\begin{align}
	S(\ell)=\frac{c}{3}\log\left[\frac{L}{\pi}\sin\frac{\pi \ell}{L}\right],
\end{align}
treating $c$ as a free parameter. In the following, we interpret the extracted $c$ as an effective measure of conformal entanglement scaling along the critical line \eqref{crit_line}, rather than as evidence for an exact CFT. The resulting $c(\beta)$ for varying system sizes is shown in Fig.~\ref{fig:centralcharge}.

\begin{figure}[h]
	\centering
	\includegraphics[width=0.6\linewidth]{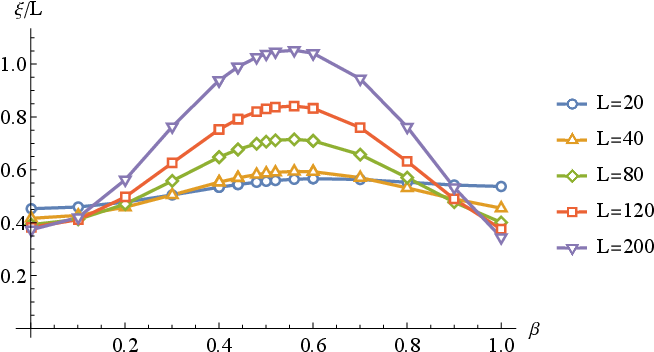}
	\caption{Relative correlation length $\xi/L$ for different system size.}
	\label{fig:corrlengthxbeta}
\end{figure}

The topological entropy scaling presented in the Calabrese-Cardy formula is expected when the system admits a low-energy conformal description with correlation length large compared to the lattice spacing. We verify this assumption by computing the spin-spin correlation
\begin{align}
	C(r)=\expval{X_{i_{0}}X_{i_{0}+r}}-\expval{X_{i_{0}}}\expval{X_{i_{0}+r}}
\end{align}
for fixed $i_{0}=L/4$, and varying $r=1, 2,  \cdots, L/2$. We fit $C(r)$ to its periodic chain form
\begin{align}
	C(r)\propto e^{-r/\xi}+e^{-\left(L-r\right)/\xi},
\end{align}
and extract the correlation length $\xi$.The resulting relative correlation length, $\xi/L$, is presented in Fig. \ref{fig:corrlengthxbeta}. The general behaviour of $\xi\propto \mathcal{O}(L)$ confirms that the line defined by \eqref{crit_line} is indeed critical.

The behavior of the central charge presented in Fig. \ref{fig:centralcharge} along is highly non-monotonic. As $\beta$ increases, the central charge is suppressed near the bulk critical point $\beta_c$, before being restored to its original, undeformed value at strong coupling. Notably, both the depth of the suppression and the sharpness of the restoration are enhanced with increasing system size, suggesting that the finite size data converges to a sharp well in the thermodynamic limit. 

The suppression of $c(\beta)$ admits a natural interpretation in light of a recent work  \cite{Verressen2025}, which argues that the bulk phase transitions proceeds through an intermediate gapless phase rather than a direct transition from a topological phase to a trivial gapped phase. In this scenario, the bulk gap closes at $\beta_c$ and extended critical modes dominate over a finite parameter range. The collapse of the central charge in this region can be understood as a boundary manifestation of this bulk gapless intermediate regime, where the assumptions underlying a simple conformal entanglement scaling break down.

Away from $\beta_{c}$, the restoration of the holographic central charge indicates the re-emergence of a stable boundary conformal description once the bulk is gapped again. Importantly, this restoration occurs even though the bulk topological entanglement entropy vanishes for $\beta>\beta_c$, demonstrating that the recovery of $c$ does not signal the reappearance of bulk topological order. Instead, it reflects the fact that both gapped phases are characterized by a finite bulk correlation length.

From this perspective, the holographic central charge is controlled by the presence or absence of long-range critical correlations in the bulk. In the vicinity of $\beta_c$, the divergence of the bulk correlation length associated with the gapless regime invalidates a simple Calabrese–Cardy scaling of the boundary entanglement entropy, leading to a strong suppression of the effective central charge extracted from finite-size data. Once the bulk gap reopens on either side of the transition and correlations become short-ranged, the boundary entanglement scaling stabilizes and the central charge is restored to its undeformed value.

\section{Final Remarks}

We studied the Castelnovo-Chamon deformation of the toric code from a symmetry and boundary approach. The deformation introduces a string tension that suppresses large loop configurations and ultimately destroys the $\mathbb{Z}_2$ topological order above the critical coupling $\beta_c$. In the bulk, this loss is sharply reflected by the vanishing of the topological entanglement entropy, while in the presence of open boundaries it is accompanied by a reorganization of its symmetry and anyon content.

Placing the model on a cylinder allowed us to track how the ordinary toric code boundary condensation and 1-form symmetry breaking are affected by the deformation. For $\beta=0$, the usual toric code picture is recovered: one 1-form symmetry remains unbroken while its anomalous partner is spontaneously broken, leading to the characteristic ground-state degeneracy. Turning on $\beta$ modifies this structure in an essential way. In particular, the deformed magnetic line operator becomes effectively non-unitary on the ground states beyond $\beta_c$ on the electric boundary. Simultaneously, the magnetic boundary (which condenses magnetic anyons for all $\beta$), also condenses electric anyons above $\beta_{c}$. From this viewpoint, the transition is not just an exchange of electric and magnetic roles (as in the $\beta=0$ boundary transition), but a genuine destruction of the symmetry and anomaly structure that stabilizes long-range entanglement in the topological phase.

We also examined the transition through a edge description, where bulk topological lines descend to global symmetries of an effective (1+1)-dimensional Hamiltonian. Along the corresponding boundary critical line, we extracted an effective central charge from DMRG data and found a striking non-monotonic dependence on $\beta$. The central charge is strongly suppressed near the bulk critical region and later returns to its undeformed value at larger $\beta$. This offers a interpretation that the boundary entanglement scaling is primarily sensitive to bulk criticality (or bulk gaplessness) rather than serving as a direct order parameter for bulk topological order.

Several directions remain open. A key problem is to better characterize the intermediate $\beta_c$, gapless regime, proposed in \cite{Verressen2025}, and match it with the symmetry and entanglement scaling diagnostics used here. It would also be valuable to track the fate of electromagnetic duality across the transition more sharply, both in the bulk and on the boundary . Finally, extending these ideas to other deformations, other gauge groups, and more generic deformations would help clarify which aspects of the ``symmetry structure degradation'' are special to this exactly solvable model and which are universal mechanisms for the destruction of topological order.

\newpage

\section{Acknowledgments}
I would like to thank Dr. Weslei Fontana for the invaluable discussions.

\bibliography{ToricCode}	
\bibliographystyle{unsrt}

\end{document}